\documentstyle[twoside,fleqn,espcrc2,epsf]{article}
\newcommand{\lsim}{\raisebox{0.3mm}{\em $\, <$} 
\hspace{-3.3mm} \raisebox{-1.8mm}{\em $\sim \,$}}

\begin{document}
\pagestyle{empty}

\title{Analysis of
the Superkamiokande atmospheric neutrino data
in the framework of four neutrino mixings
\thanks{Talk presented at ``nufact00'', Monterey, CA, USA
		May 22-26, 2000.}}

\author{Osamu Yasuda\\
        {\ }\\
        Department of Physics, Tokyo Metropolitan University\\
        1-1 Minami-Osawa Hachioji, Tokyo 192-0397, Japan}

\begin{abstract}
Superkamiokande atmospheric neutrino data
for 990 days are analyzed in
the framework of four neutrinos
without imposing constraints of Big Bang Nucleosynthesis.  It
is shown that the wide range of the oscillation parameters is allowed
at 90\% confidence level ($0.1\lsim |U_{s1}|^2+|U_{s2}|^2\le 1$).
\end{abstract}

\maketitle


\section{Introduction}
It has been known that
three different kinds of experiments suggest neutrino oscillations:
the solar neutrino deficit
\cite{solar}
the atmospheric neutrino anomaly
\cite{atm,toshito}
and the LSND data \cite{lsnd}.  If one assumes that all these three are
caused by neutrino oscillations then one needs at least four species of
neutrinos.
Furtherover, it has been shown \cite{oy,bggs} that the $4\times 4$ MNS
matrix splits approximately into two $2\times 2$ block diagonal
matrices if one imposes \cite{goswami} the constraint of
the reactor data \cite{bugey} and
if one demands that
the number $N_\nu$ of effective neutrinos 
in Big Bang Nucleosynthesis (BBN) be less than four.
In this case the solar neutrino deficit is explained by
$\nu_e\leftrightarrow\nu_s$ oscillations with the Small Mixing Angle
(SMA) MSW solution and the atmospheric neutrino anomaly is
accounted for by $\nu_\mu\leftrightarrow\nu_\tau$.  On the other hand,
some people have given conservative estimate for $N_\nu$ \cite{he4}
and if their estimate is correct then the
constraints on the mixing angles of sterile neutrinos are not
as strongs as in the case of $N_\nu<4$ and the conditions
one has to take into account are the data of the reactor,
the solar neutrinos and the atmospheric neutrinos. Recently
Giunti, Gonzalez-Garcia and Pe\~na-Garay \cite{ggp} have analyzed the
solar neutrino data in the four neutrino scheme without BBN
constraints.  They have shown that the scheme is reduced to the two
neutrino framework in which only one free parameter
$c_s\equiv|U_{s1}|^2+|U_{s2}|^2$ appears in the analysis.  Their
conclusion is that the SMA MSW solution exists for the entire region
of $0\le c_s
\le 1$, while the Large Mixing Angle (LMA) and Vacuum Oscillation (VO)
solutions survive only for $0\le c_s \lsim 0.2$ and $0\le c_s \lsim
0.4$, respectively.
In this talk I will discuss the analysis of
the Superkamiokande atmospheric neutrino data for 990 days \cite{toshito}
(contained and upward going through $\mu$ events)
in the same scheme as in \cite{ggp}, i.e., in the four neutrino scheme
with all the constraints of accelerators and reactors but without
BBN constraints.
Details and more references are given in \cite{y}.

\section{Four neutrino scheme}
\begin{figure}[h]
\vglue 0.5cm
\hglue -2.5cm
\hbox to\hsize{\hss\epsfxsize=6.2cm\epsfbox{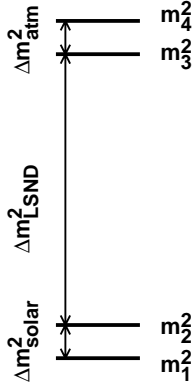}\hss}
\vglue -1.5cm
\caption{\label{fig:pattern}Four neutrino scheme}
\end{figure}

Here I adopt the notation in \cite{oy} for the $4\times 4$ MNS matrix:
\begin{eqnarray}
\left( \begin{array}{c} \nu_e  \\ \nu_{\mu} \\ 
\nu_{\tau}\\\nu_s \end{array} \right)
=\left(
\begin{array}{cccc}
U_{e1} & U_{e2} &  U_{e3} &  U_{e4}\\
U_{\mu 1} & U_{\mu 2} & U_{\mu 3} & U_{\mu 4}\\
U_{\tau 1} & U_{\tau 2} & U_{\tau 3} & U_{\tau 4}\\
U_{s1} & U_{s2} &  U_{s3} &  U_{s4}
\end{array}\right)
\left( \begin{array}{c} \nu_1  \\ \nu_2 \\ 
\nu_3\\\nu_4 \end{array} \right).
\nonumber\\
\label{eqn:u}
\end{eqnarray}
I can assume without loss of generality that
$m_1^2$ $<$ $m_2^2$ $<$ $m_3^2$ $<$ $m_4^2$.
Three mass scales
$\Delta m_\odot^2$ $\sim$ ${\cal O}(10^{-5}{\rm eV}^2)$ or
${\cal O}(10^{-10}{\rm eV}^2)$,
$\Delta m_{\rm atm}^2$ $\sim$ ${\cal O}(10^{-2}{\rm eV}^2)$,
$\Delta m_{\rm LSND}^2$ $\sim$ ${\cal O}(1{\rm eV}^2)$
are necessary to explain the data of the solar
neutrinos, the atmospheric neutrinos and LSND,
so I assume that three independent mass squared
differences
are $\Delta
m_\odot^2$, $\Delta m_{\rm atm}^2$, $\Delta m_{\rm LSND}^2$.  It has
been known \cite{oy,bgg} that schemes with three
degenerate masses and one distinct massive state do not work
to account for all the three neutrino anomalies, but
schemes with two degenerate massive states
($m_{1}^2 \simeq m_{2}^2 \ll m_{3}^2 \simeq m_{4}^2$,
where (a) $(\Delta m_{21}^2, \Delta m_{43}^2)=
(\Delta m_\odot^2,\Delta m_{\rm atm}^2)$ or (b)
$(\Delta m_{21}^2, \Delta m_{43}^2)=
(\Delta m_{\rm atm}^2,\Delta m_\odot^2)$)
do.  As far as the analyses of atmospheric
neutrinos and solar neutrinos are concerned,
the two cases (a) and (b) can be treated in the same manner,
so I assume $\Delta m_{21}^2=\Delta m_\odot^2$,
$\Delta m_{43}^2=\Delta m_{\rm atm}^2$ in the following
(See Fig.\ref{fig:pattern}).

For the range of the $\Delta m^{2}$
suggested by the LSND data, which is given by 0.2
eV$^2~\lsim \Delta m^{2}_{\mbox{\rm{\scriptsize LSND}}}
\lsim$ 2 eV$^2$ when
combined with the data of Bugey \cite{bugey} and E776 \cite{e776},
the constraint by the Bugey data is very stringent and
\begin{eqnarray}
|U_{e3}|^2+|U_{e4}|^2\lsim 10^{-2}
\end{eqnarray}
has to be satisfied \cite{goswami,oy,bgg}.  Therefore I put
$U_{e3}=U_{e4}=0$ for simplicity in the following discussions.  Also
in the analysis of atmospheric neutrinos, $|\Delta m_\odot^2L/4E|\ll
1$ is satisfied for typical values of the neutrino path length $L$ and
the neutrino energy $E$, so I assume $\Delta m_{21}^2= 0$ for
simplicity throughout this talk.

Having assumed $U_{e3}=U_{e4}=0$ and $\Delta m_{21}^2= 0$,
I have only mixings among $\nu_\mu$, $\nu_\tau$, $\nu_s$
in the analysis of atmospheric neutrinos, and the
Schr\"odinger equation I have to consider is
\begin{eqnarray}
&{\ }&i {d \over dx} \left( \begin{array}{c} \nu_{\mu}(x) \\ 
\nu_{\tau}(x) \\ \nu_s (x)
\end{array} \right)={\cal M}\left( \begin{array}{c}
\nu_{\mu}(x) \\ \nu_{\tau}(x) \\ \nu_s (x)
\end{array} \right),\nonumber\\
{\cal M}&\equiv& 
\left[ \widetilde U {\rm diag} \left(-\Delta E_{32} ,0,\Delta E_{43}
\right) \widetilde U^{-1}\right.\nonumber\\
&{\ }&\left.+{\rm diag} \left(0,0,A(x) \right) \right]
\label{eqn:sch}
\end{eqnarray}
where $\Delta E_{ij}\equiv\Delta m_{ij}^2/2E$,
$A(x)\equiv G_F N_n(x)/\sqrt{2}$ stands for the effect due to the
neutral current interactions between $\nu_\mu$, $\nu_\tau$ and matter in the
Earth and
\begin{eqnarray}
\widetilde U&\equiv&\left(
\begin{array}{ccc}
 U_{\mu 2} & U_{\mu 3}&U_{\mu 4} \\
 U_{\tau 2} & U_{\tau 3}&U_{\tau 4} \\
 U_{s2} &  U_{s3}&U_{s4} 
\end{array}\right)\nonumber\\
&=&e^{i({\pi \over 2}-\theta_{34})\lambda_7}
D^{-1} e^{i\theta_{24}\lambda_5} D~
e^{i(\theta_{23}-{\pi \over 2})\lambda_2},
\label{eqn:mns3}
\end{eqnarray}
with $D\equiv{\rm diag}\left(e^{i\delta_1/2},1,e^{-i\delta_1/2} \right)$
($\lambda_j$ are the $3\times 3$ Gell-Mann matrices)
is the reduced $3\times 3$ MNS matrix.
This MNS matrix $\widetilde U$
is obtained by substitution
$\theta_{12}\rightarrow\theta_{23}-\pi/2$,
$\theta_{13}\rightarrow\theta_{24}$,
$\theta_{12}\rightarrow\pi/2-\theta_{34}$,
$\delta\rightarrow\delta_1$
in the standard parametrization in \cite{pdg}.
Since $\nu_e$ does not oscillate with any other
neutrinos, the only oscillation probability which is required in the analysis
of atmospheric neutrinos is $P(\nu_\mu\rightarrow\nu_\mu)$.

In the following analysis I will consider the situation where
non-negligible contribution from the largest mass squared difference
$\Delta m^2_{32}$ appears in the
oscillation probability $P(\nu_\mu\rightarrow\nu_\mu)$.
In short baseline experiments where
$|\Delta E_{21}|$ and $|\Delta E_{43}|$ can be neglected,
the disappearing probability $P(\nu_\mu\rightarrow\nu_\mu)$
is given by
\begin{eqnarray}
P(\nu_\mu\rightarrow\nu_\mu)&=&1-4|U_{\mu 2}|^2(1-|U_{\mu 2}|^2)
\nonumber\\
&{\ }&\times
\sin^2\left({\Delta m_{32}^2L \over 4E}\right)
\end{eqnarray}
so that the mixing angle $\sin^22\theta_{\mbox{\rm{\scriptsize SBL}}}$
is given by
\begin{eqnarray}
\sin^22\theta_{\mbox{\rm{\scriptsize SBL}}}=4c_{24}^2s_{23}^2
\left(1-c_{24}^2s_{23}^2\right).
\end{eqnarray}
It turns out in the final results that
$\sin^22\theta_{\mbox{\rm{\scriptsize SBL}}}$
can be as large as 0.5 in the allowed region of the atmospheric
neutrino data.
To avoid contradiction with the negative result
of the CDHSW disappearing experiment on
$\nu_\mu\rightarrow\nu_\mu$ \cite{cdhsw}, I will
take $\Delta m^{2}_{32}$=0.3eV$^2$
as a reference value.

\begin{figure}[p]
\vglue -2.9cm\hglue -3.0cm
\hbox to\hsize{\hss\epsfxsize=8.2cm\epsfbox{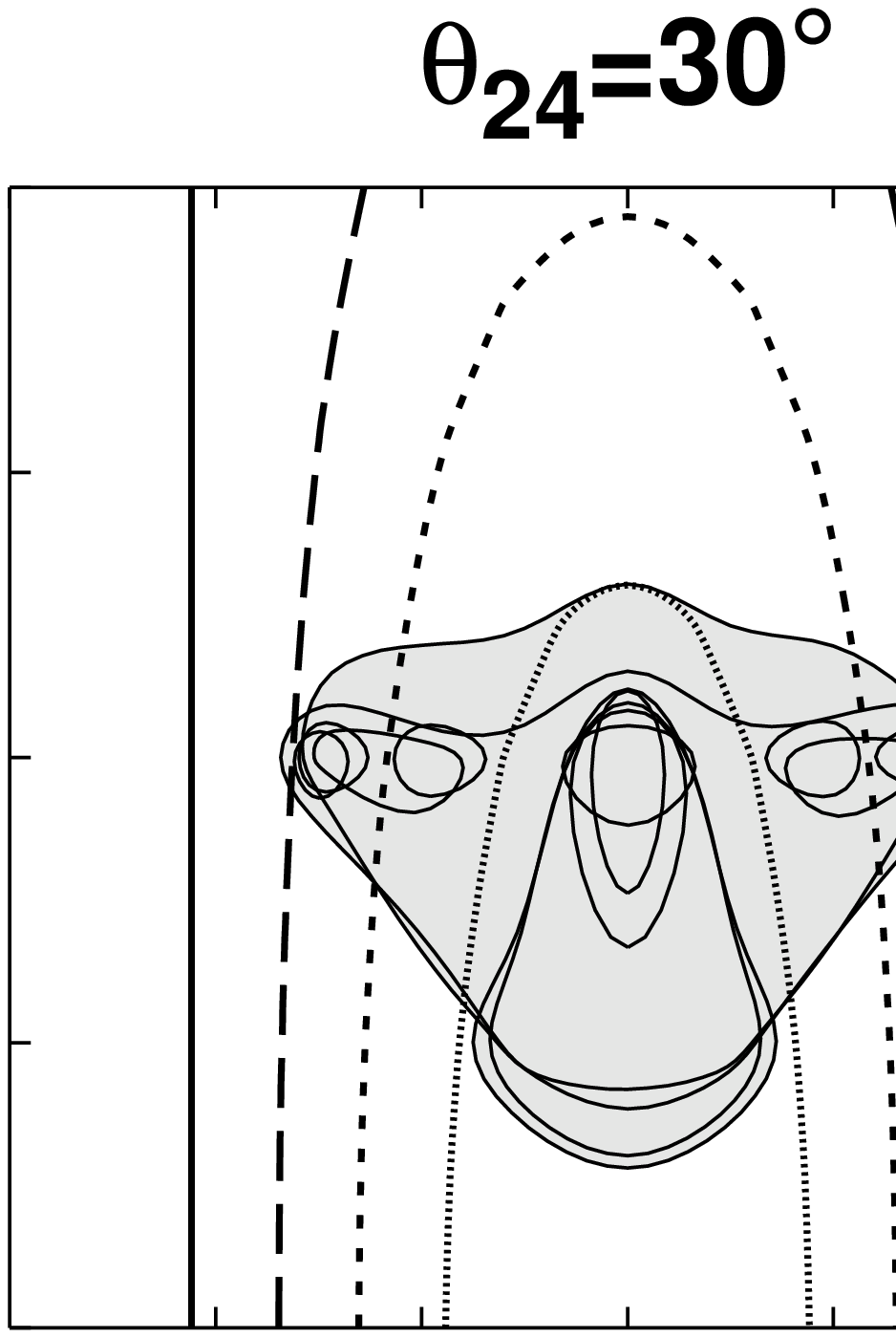}\hss}
\vglue -8.25cm\hglue 2.9cm
\hbox to\hsize{\hss\epsfxsize=8.2cm\epsfbox{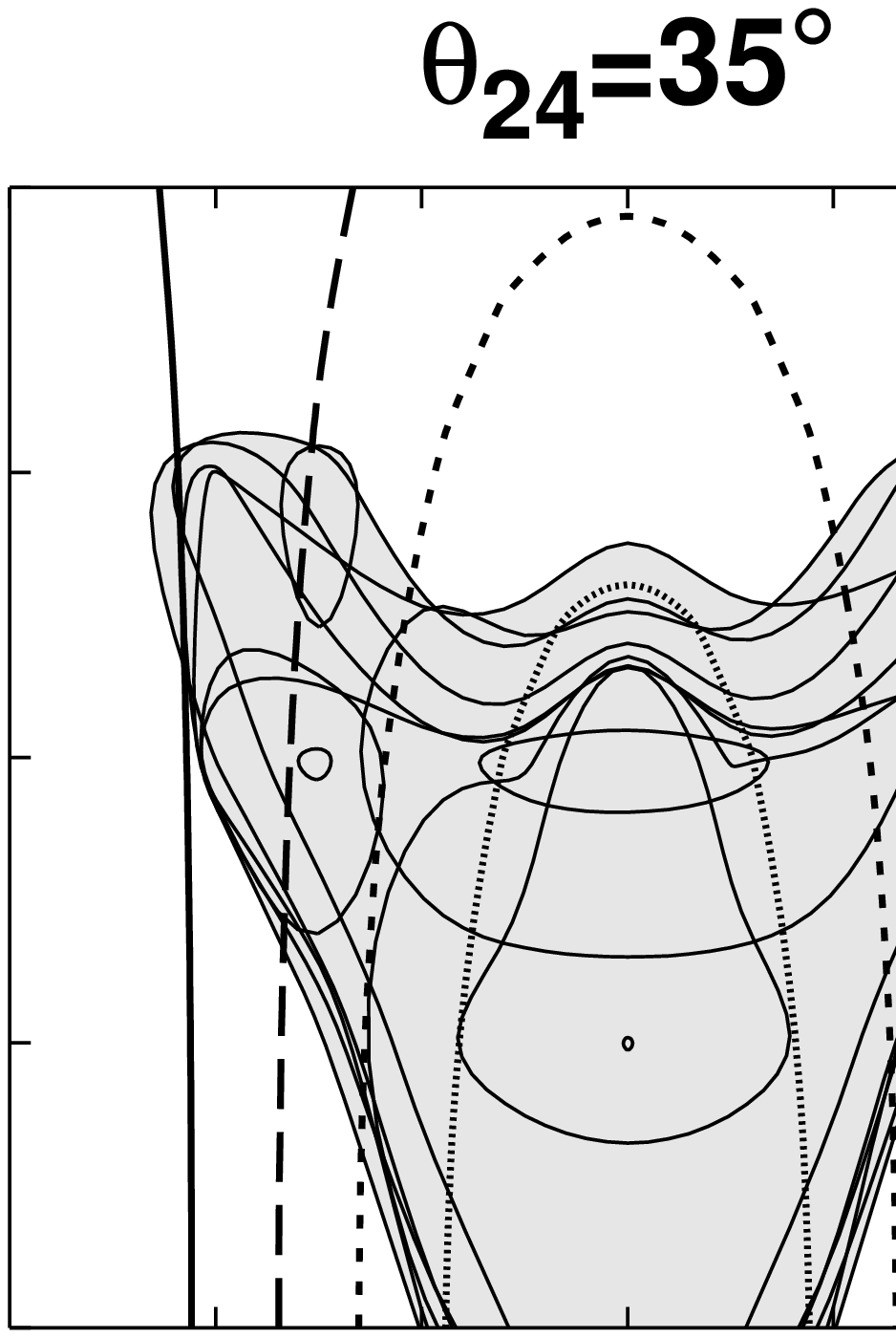}\hss}
\vglue -2.9cm\hglue -3.0cm
\hbox to\hsize{\hss\epsfxsize=8.2cm\epsfbox{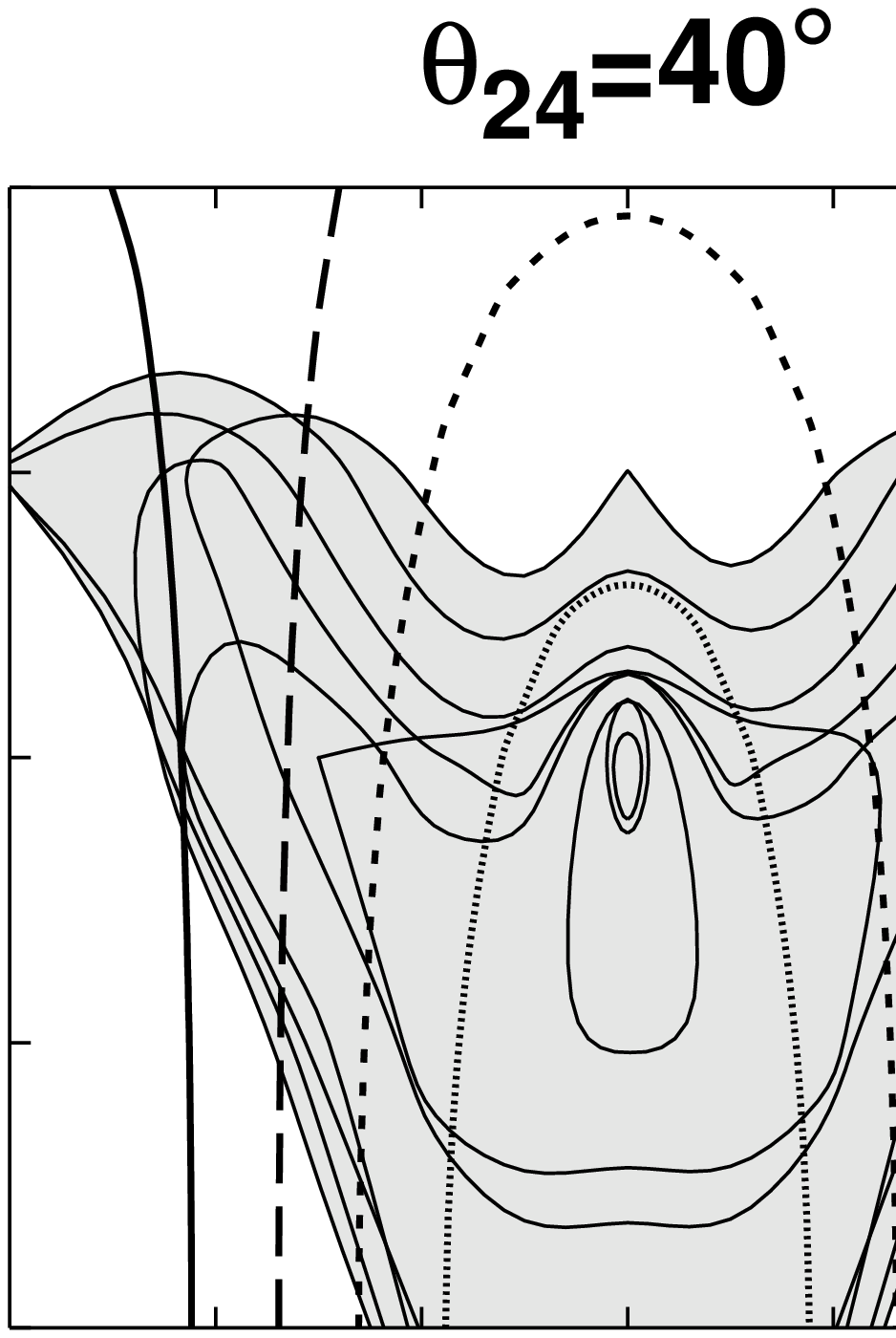}\hss}
\vglue -8.25cm\hglue 2.9cm
\hbox to\hsize{\hss\epsfxsize=8.2cm\epsfbox{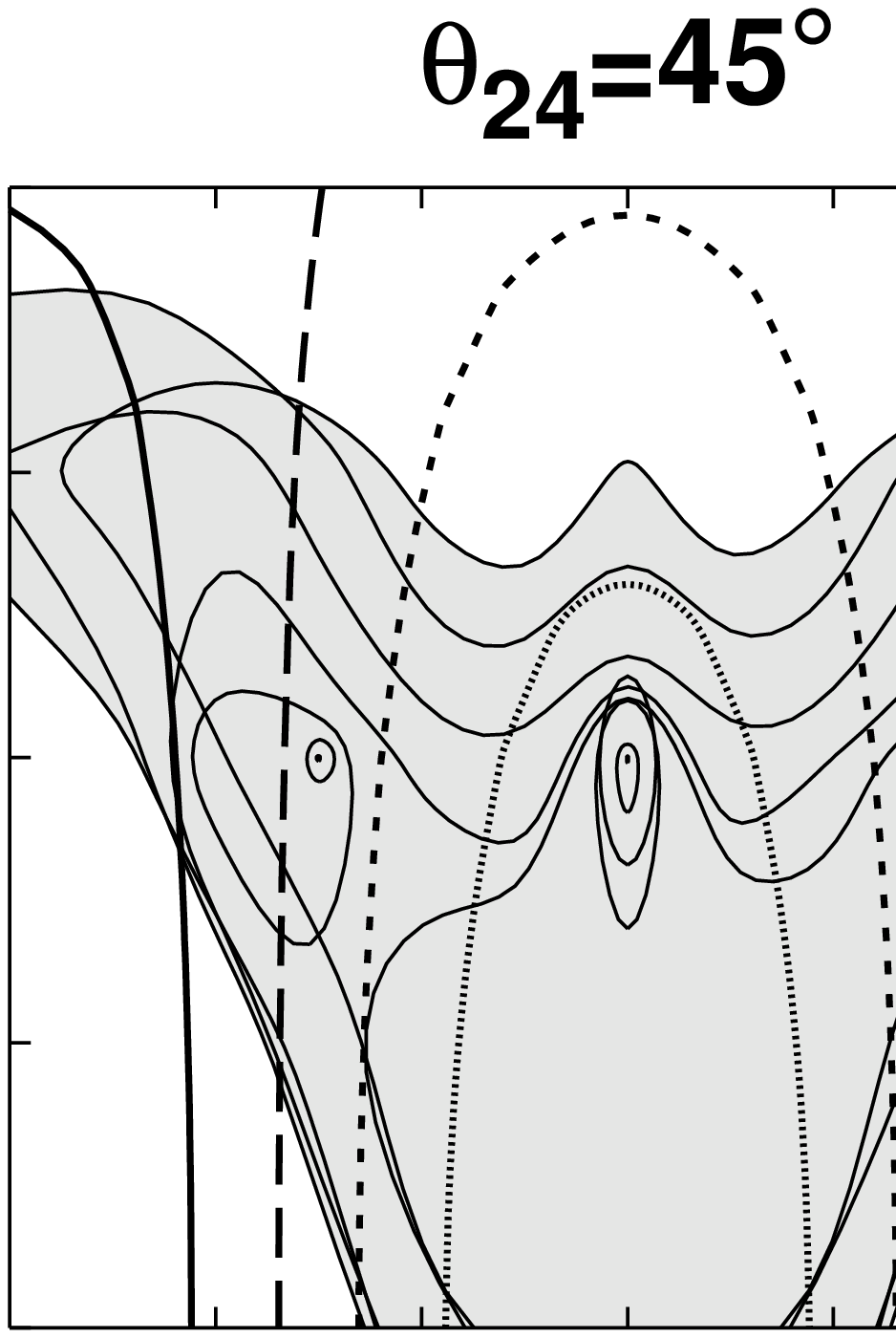}\hss}
\vglue -2.9cm\hglue -3.0cm
\hbox to\hsize{\hss\epsfxsize=8.2cm\epsfbox{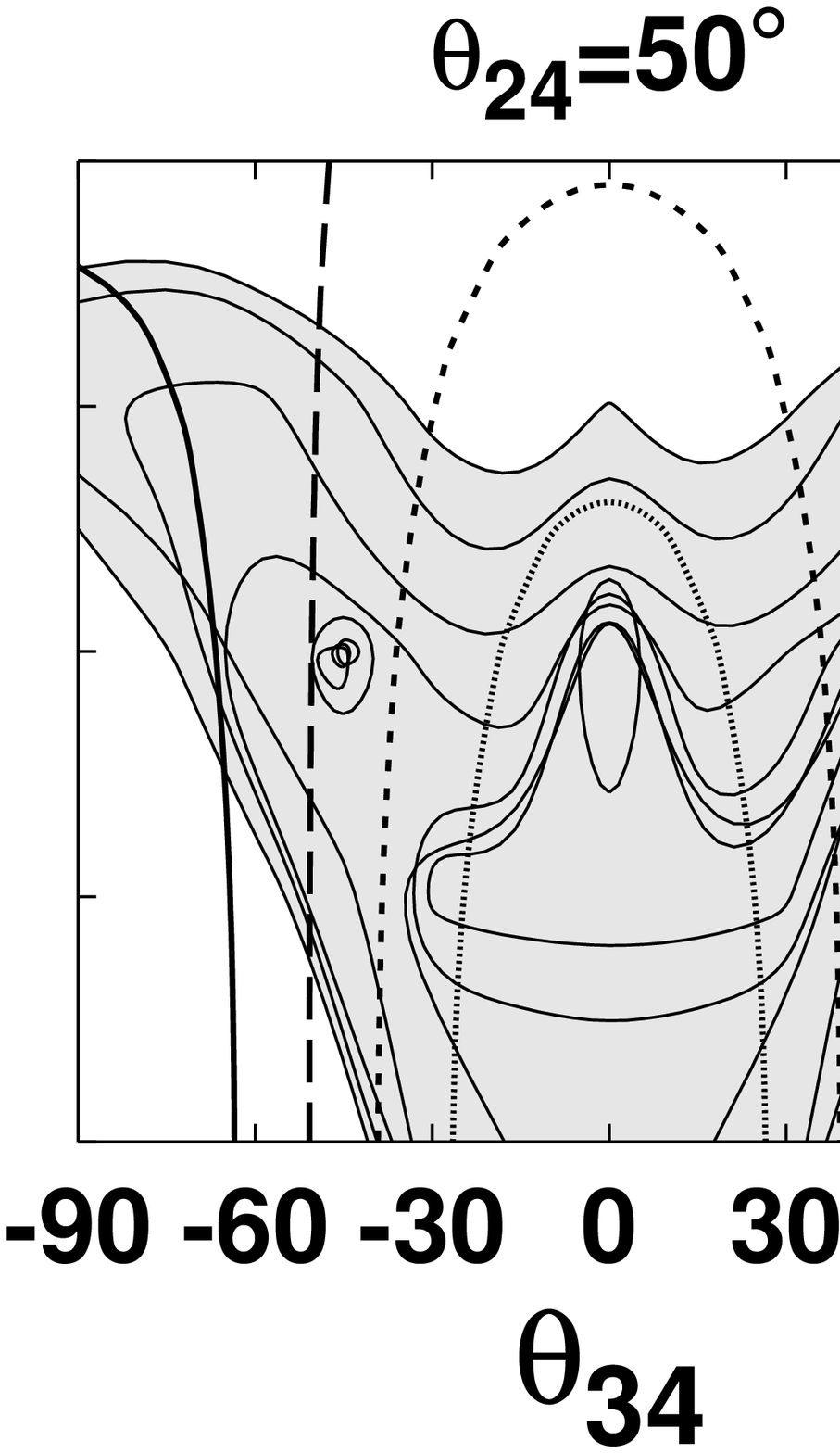}\hss}
\vglue -8.25cm\hglue 2.9cm
\hbox to\hsize{\hss\epsfxsize=8.2cm\epsfbox{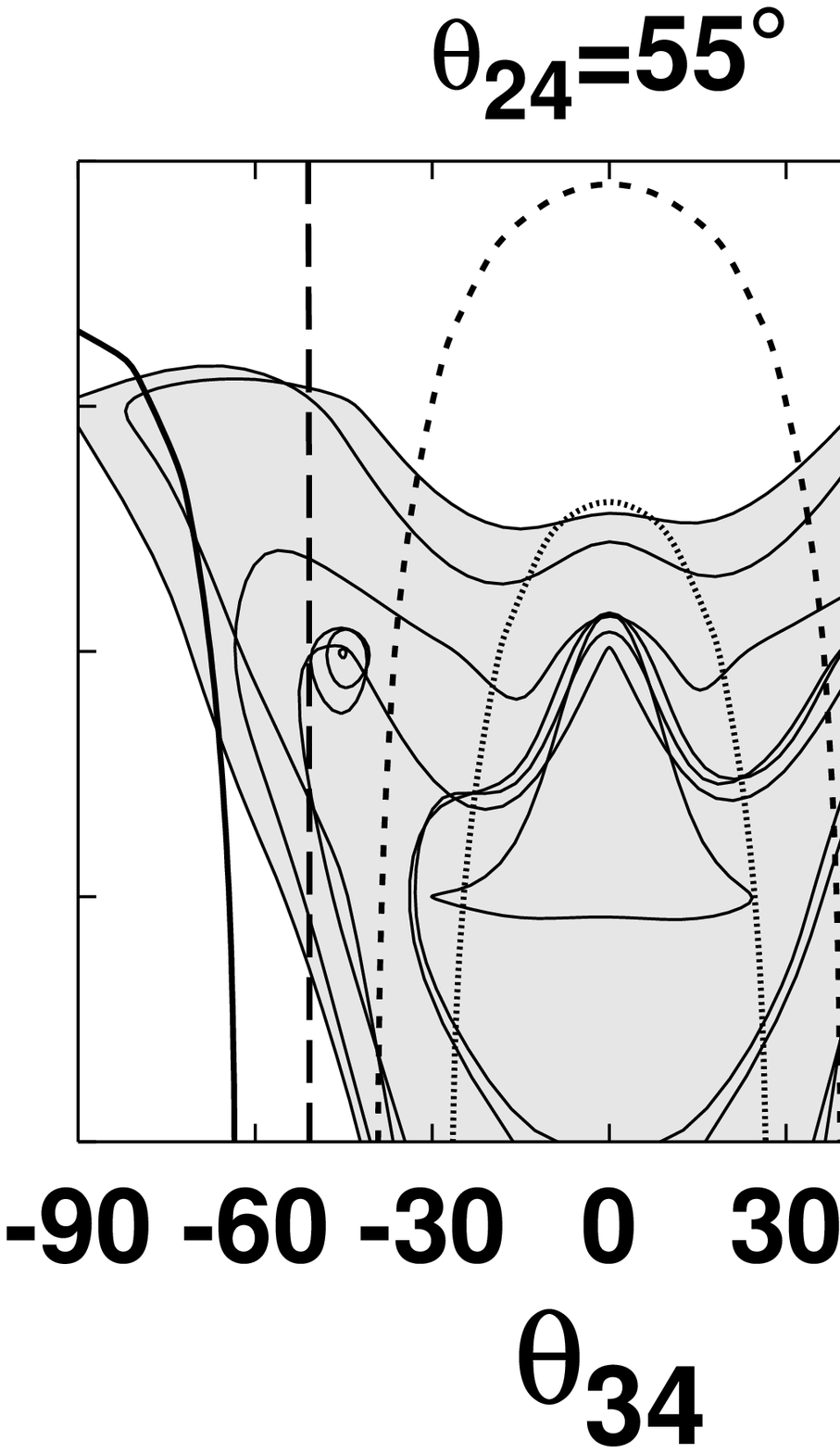}\hss}
\vglue -1.75cm\hglue 0.9cm
\hbox to\hsize{\hss\epsfxsize=8.2cm\epsfbox{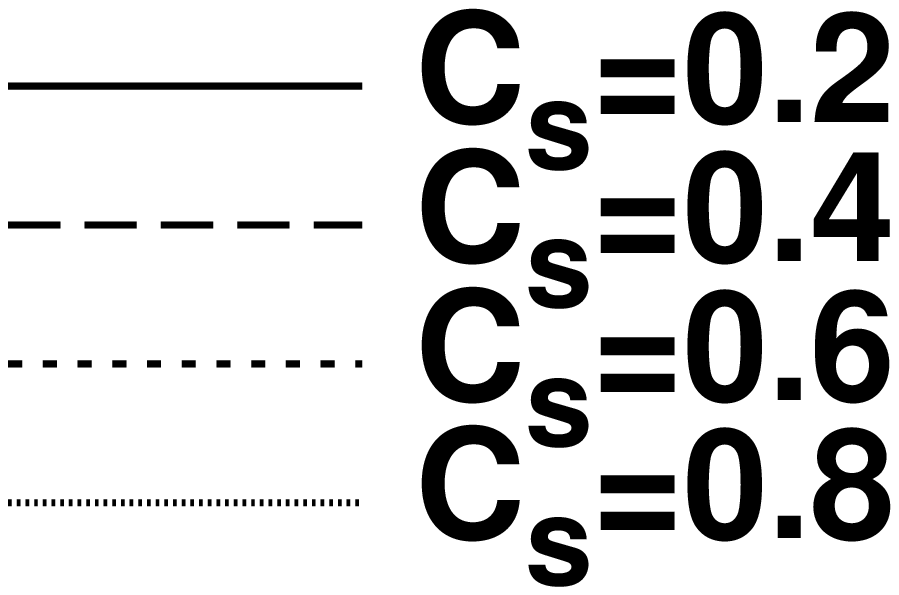}\hss}
\vglue -6.0cm
\caption{\label{fig:allowed}Allowed region at 90\%CL for
$\delta_1=\pi/2$.}
\begin{minipage}{16cm}
The shadowed area is the
allowed region projected on the $(\theta_{34},~
\theta_{23})$ plane for various values of $\Delta m_{43}^2$
($10^{-3.5}$eV$^2\le\Delta m_{43}^2\le 10^{-2}$eV$^2$)
for each value of $\theta_{24}=30^\circ,
\cdots,55^\circ$, respectively,
and the thin solid lines are boundary of the allowed region
for various values of $\Delta m_{43}^2$.
The solid, dashed, coarse dotted and fine dotted lines
stand for the contours of
$c_s\equiv|U_{s1}|^2+|U_{s2}|^2$
=$|c_{23}c_{34}+s_{23}s_{24}s_{34}e^{i\delta_1}|^2=0.2,
0.4, 0.6, 0.8$,
respectively.  Solutions with $c_s\lsim 0.2$ exist
for $40^\circ\lsim\theta_{24}\lsim 55^\circ$
and they can have Large Mixing Angle solutions of the solar
neutrino problem.
\end{minipage}
\end{figure}

\newpage
{\ }
\newpage
\section{Analysis of the atmospheric neutrino data}
\begin{figure}[h]
\vglue -1.0cm
\hbox to\hsize{\hss\epsfxsize=4.2cm\epsfbox{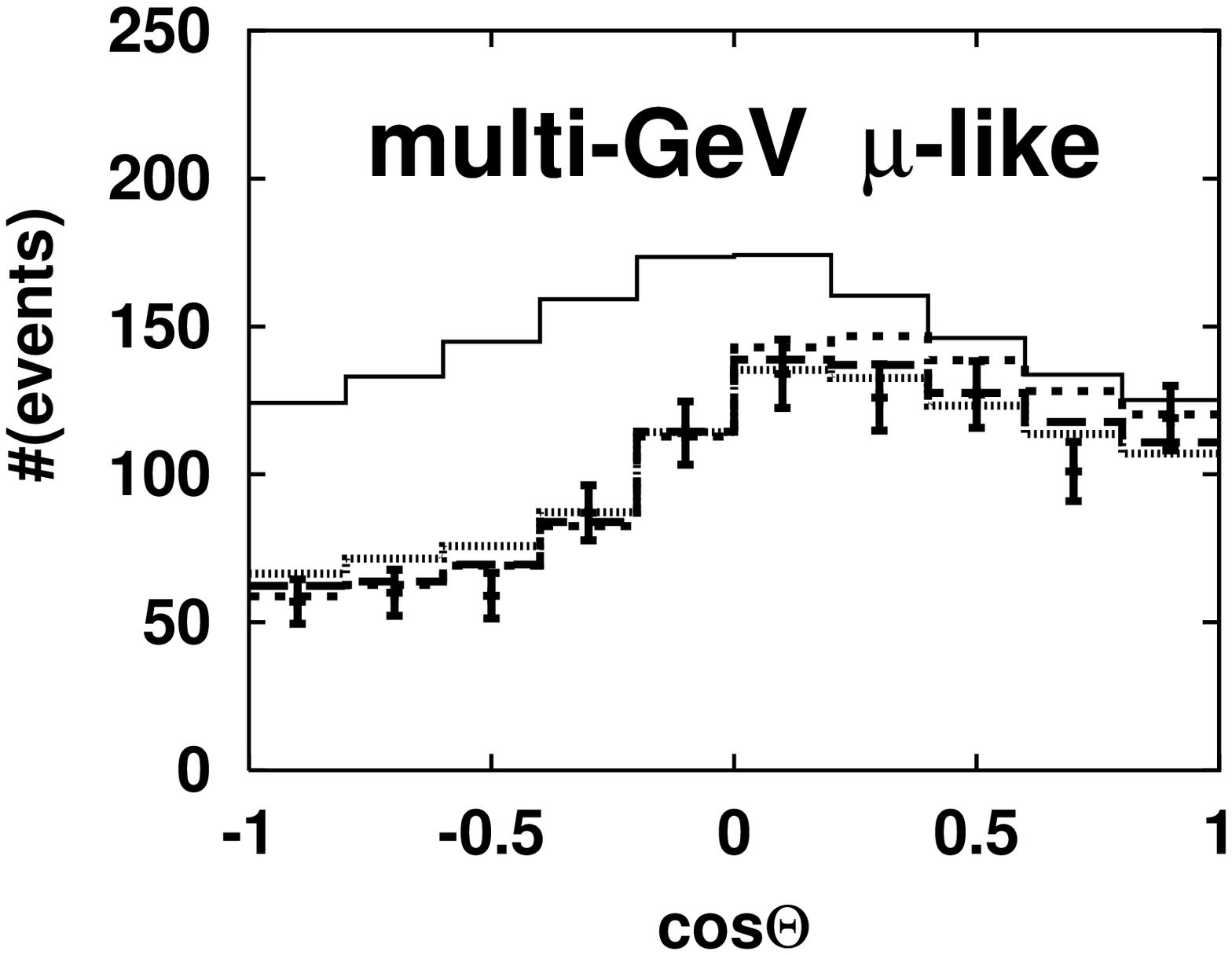}\hss}
\vglue 0.9cm
\hbox to\hsize{\hss\epsfxsize=4.2cm\epsfbox{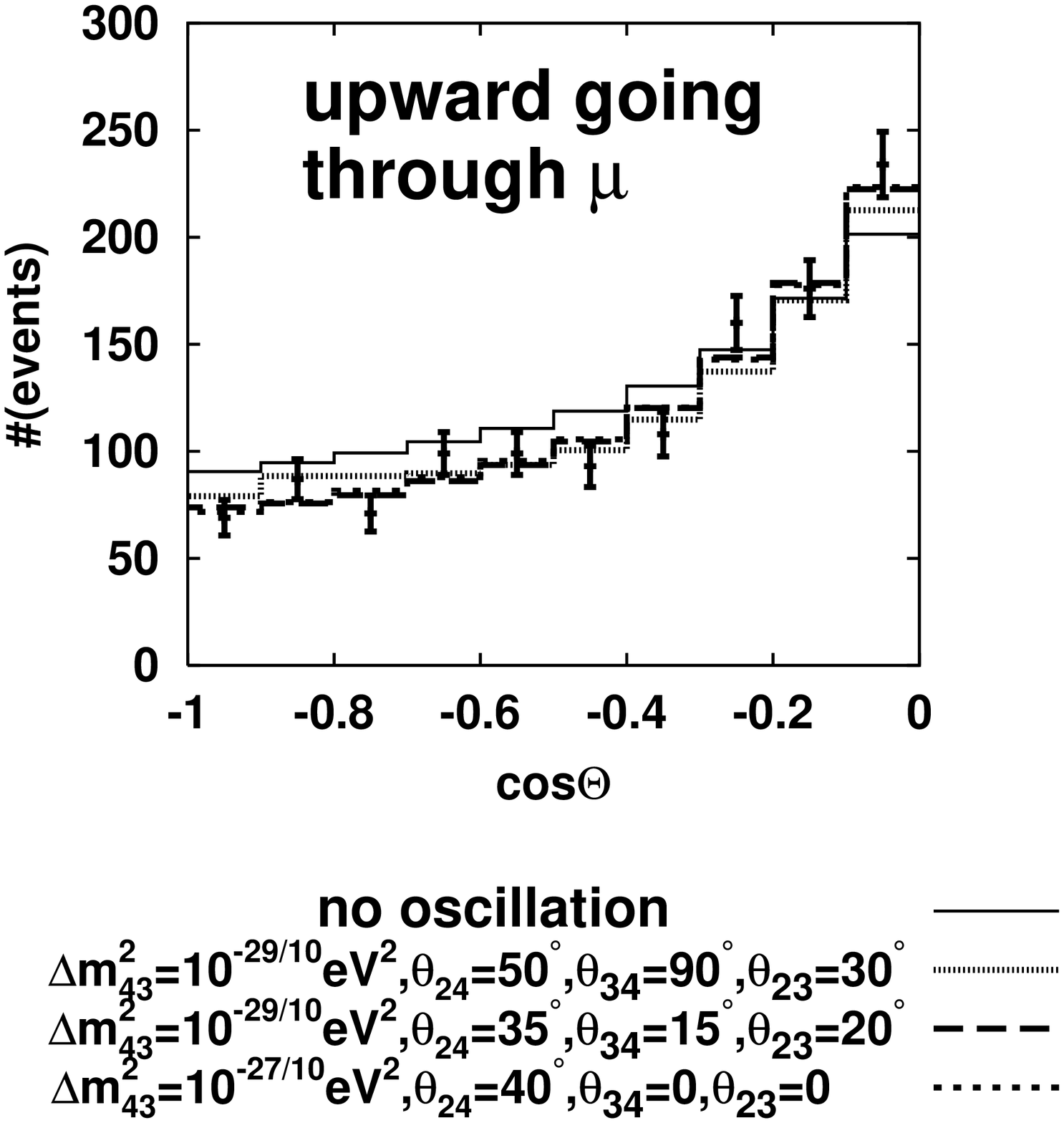}\hss}
\vglue -0.5cm
\caption{\label{fig:zenith}Zenith angle dependence}
\end{figure}

I calculate the disappearance probability
$P(\nu_\mu\rightarrow\nu_\mu)$ by solving (\ref{eqn:sch})
numerically, and evaluate the number of events.
Then I define $\chi^2$ as
\begin{eqnarray}
\chi^2=\chi_{\rm sub-GeV}^2+\chi_{\rm multi-GeV}^2
+\chi_{\rm through}^2
\label{eqn:chi}
\end{eqnarray}
where $\chi_{\rm sub-GeV}^2$, $\chi_{\rm multi-GeV}^2$
and $\chi_{\rm through}^2$
are $\chi^2$ for sub-GeV, multi-GeV, and upward going through
$\mu$ events, respectively.
I have evaluated $\chi^2$ for $\theta_{24}= (25+5j)^\circ$
($j=0,\cdots,7$), $\theta_{34}= 15j^\circ$ ($j=-6,\cdots,6$),
$\theta_{23}= 10j^\circ$ ($j=0,\cdots,4$),
$\delta_1 = 0^\circ, 45^\circ, 90^\circ$,
$\Delta m^2_{43}= 10^{-4+j/10}$eV$^2$ ($j=5,\cdots,20$)
and it is found that $\chi^2$ has the minimum value
$\chi^2_{\rm min}=43.1$ $(\chi_{\rm sub-GeV}^2=19.0$,
$\chi_{\rm multi-GeV}^2=13.2$,
$\chi_{\rm through}^2=11.6$)
for $\Delta m_{43}^2=10^{-2.9}{\rm eV}^2=
1.3\times 10^{-3}{\rm eV}^2$,
$(\theta_{24},\theta_{34},\theta_{23})=
(35^\circ,15^\circ,20^\circ)$,
$\delta_1=0$ with 45 degrees of freedom.
For pure $\nu_\mu\leftrightarrow\nu_\tau$
($\theta_{34}=\theta_{23}=0$),
the best fit is obtained
$\chi^2_{\rm min}(\nu_\mu\leftrightarrow\nu_\tau)=48.3$,
$(\chi_{\rm sub-GeV}^2=19.8,
\chi_{\rm multi-GeV}^2=17.0,
\chi_{\rm through}^2=10.6)$
for
$\Delta m_{43}^2=
2.0\times 10^{-3}{\rm eV}^2$,
$(\theta_{24},\theta_{34},\theta_{23})=
(40^\circ,0^\circ,0^\circ)$.
The allowed regions at 90\%CL are obtained by
$\chi^2\le \chi^2_{\rm min}+\Delta \chi^2$,
where $\Delta \chi^2$=9.2 for five degrees
of freedom.
In Fig.\ref{fig:allowed} the
allowed region at 90\% confidence level is depicted as a shadowed
area in the $(\theta_{34},~
\theta_{23})$ plane for various values of $\Delta m_{43}^2$
($10^{-3.5}$eV$^2\le\Delta m_{43}^2\le 10^{-2}$eV$^2$)
for each value of $\theta_{24}=30^\circ,
\cdots,55^\circ$ and for $\delta_1=\pi/2$, together with
lines $c_s$=constant.  A few remarks are in order.
(1)~Pure $\nu_\mu\leftrightarrow\nu_s$ oscillation,
which is given by $\theta_{34}=\pm 90^\circ$,~
$\theta_{23}=0$, is excluded at 99.7\%CL for any
value of $\Delta m_{43}^2$, $\theta_{24}$, $\delta_1$
and this is consistent with the claim \cite{toshito}
by the Superkamiokande group.
(2)~For generic value of $(\theta_{34},~\theta_{23})$,
the oscillation is hybrid not only with
$\nu_\mu\leftrightarrow\nu_\tau$ and
$\nu_\mu\leftrightarrow\nu_s$ but also with
$\Delta m_{43}^2$ and $\Delta m_{32}^2$.
To illustrate this, let me give the expressions of
oscillation probability in vacuum in the case of
$\delta_1=\pi/2$:
\begin{eqnarray}
P(\nu_\mu\rightarrow\nu_\tau)&=&2c_{24}^2s_{23}^2
(c_{23}^2s_{34}^2+s_{23}^2s_{24}^2c_{34}^2)\nonumber\\
&+&c_{23}^2c_{34}^2\sin^22\theta_{24}
\sin^2\left(
{\Delta m_{43}^2L \over 4E}\right)\nonumber\\
P(\nu_\mu\rightarrow\nu_s)&=&2c_{24}^2s_{23}^2
(c_{23}^2c_{34}^2+s_{23}^2s_{24}^2s_{34}^2)\nonumber\\
&+&c_{23}^2s_{34}^2\sin^22\theta_{24}
\sin^2\left(
{\Delta m_{43}^2L \over 4E}\right)\nonumber\\
P(\nu_\mu\rightarrow\nu_\mu)&=&1-2c_{24}^2s_{23}^2
\left(1-c_{24}^2s_{23}^2\right)-c_{23}^2
\nonumber\\
&\times&\sin^22\theta_{24}
\sin^2\left(
{\Delta m_{43}^2L \over 4E}\right),
\label{eqn:prob}
\end{eqnarray}
where I have averaged over rapid oscillations due to
$\Delta m_{32}^2$.
As is seen in (\ref{eqn:prob}), roughly speaking,
$\theta_{34}$ represents the ratio of
$\nu_\mu\leftrightarrow\nu_\tau$ and
$\nu_\mu\leftrightarrow\nu_s$, whereas
$\theta_{23}$ indicates the contribution of
$\sin^2(\Delta m_{32}^2L/4E)$ in oscillations.
Zenith angle dependence of the
$\mu$-like multi-GeV events and the upward going through $\mu$
events are shown in Fig.\ref{fig:zenith} for a few sets of
the oscillation parameters.
The disappearance probability behaves like
$1-P(\nu_\mu\rightarrow\nu_\mu)=\alpha+\beta
\sin^2(\Delta m_{43}^2L/4E)$ ($\alpha$, $\beta$ are
constant) and $\alpha>0$ is satisfied whenever $\theta_{23}\ne0$.
Because of this constant $\alpha$, which never appears
in the analysis of the two flavor framework, the fit for $\theta_{23}\ne0$
tends to be better than in the case of $\theta_{23}=0$.
The reason why the best fit point is
slightly away from pure $\nu_\mu\leftrightarrow\nu_\tau$
case and the reason why an exotic solution like
($\theta_{24}$, $\theta_{23}$, $\theta_{34}$)
=($45^\circ$, $30^\circ$, $90^\circ$), $\delta_1=90^\circ$,
$\Delta m_{43}^2$=1.3$\times 10^{-3}$eV$^2$
is allowed is because a better fit to the multi-GeV contained events
compensates a worse fit to
the upward going through $\mu$ events, and in total the case of hybrid
oscillations fits better to the data
(Notice that the fit of $\nu_\mu\leftrightarrow\nu_s$ scenario
to the contained events is known to be good \cite{fvy,gnpv}
and in the present case the fit becomes even better
due to the presence of $\alpha$).

\section{Conclusions}

I have shown in the framework of four neutrino oscillations
without assuming the BBN constraints
that the Superkamiokande atmospheric neutrino data
are explained by wide range of the oscillation parameters
which implies hybrid oscillations with $\nu_\mu\leftrightarrow\nu_\tau$
and $\nu_\mu\leftrightarrow\nu_s$ as well as
with $\Delta m_{\mbox{\rm\scriptsize atm}}^2$ and
$\Delta m_{\mbox{\rm\scriptsize LSND}}^2$.
The case of pure $\nu_\mu\leftrightarrow\nu_s$ is excluded
at 3.0$\sigma$CL in good agreement with the Superkamiokande
analysis.  It is found by combining the analysis on the solar
neutrino data by Giunti, Gonzalez-Garcia and Pe\~na-Garay
that the LMA and VO solutions
as well as SMA solution of the solar neutrino problem
are allowed.


\section*{Acknowledgments}

This
research was supported in part by a Grant-in-Aid for Scientific
Research of the Ministry of Education, Science and Culture,
\#12047222, \#10640280.



\end{document}